# Effect of the indium addition on the superconducting property and the impurity phase in polycrystalline $SmFeAsO_{1-x}F_x$

Masaya Fujioka,[1] Toshinori Ozaki,[1] Hiroyuki Takeya,[1] Keita Deguchi,[1,2] Satoshi Demura,[1,2] Hiroshi Hara,[1,2] Tohru Watanabe,[1,2] Denholme Saleem James,[1] Hiroyuki Okazaki,[1] Takahide Yamaguchi,[1] Hiroaki Kumakura[1] and Yoshihiko Takano[1,2]

[1] *National Institute for Materials Science, 1-2-1 Sengen, Tsukuba, Ibaraki 305-0047, Japan*
[2] *University of Tsukuba, 1-1-1Tennodai, Tsukuba, Ibaraki 305-0001, Japan*

(Abstract)

We report enhancement in the magnetic critical current density of indium added polycrystalline $SmFeAsO_{1-x}F_x$. The value of magnetic $J_c$ is around $2.5 \times 10^4$ A/cm$^2$ at 4.2 K under self-magnetic field. Polycrystalline $SmFeAsO_{1-x}F_x$ is mainly composed of the superconducting grains and a little of amorphous FeAs compounds. These areas randomly co-exist and amorphous areas are located between superconducting grains. Therefore, the superconducting current is prevented by the amorphous areas. In this study, it is found that indium addition to polycrystalline $SmFeAsO_{1-x}F_x$ removes these amorphous areas and induces the bringing together the superconducting grains. It means the total contact surfaces of grains are increased. We suggest that the enhancement of the magnetic critical current density is a direct effect of the indium addition.

INTRODUCTION

Many rare-earth iron oxypnictides (Re-1111) [1-10] have been readily explored since the discovery of superconductivity in LaFeAsO$_{1-x}$F$_x$ (La-1111) [1] with a superconducting transition of 26 K in February 2008. In this 1111 system, a transition temperature ($T_c$) above 50 K was reported when La was replaced by Pr, Nd, Sm and Gd. Moreover, an upper critical magnetic field ($H_{c2}$) over 100 T was deduced for Sm-1111 and Nd-1111 [6]. However, polycrystalline samples of 1111 system show the significant depression of critical current density ($J_c$) due to the grain boundary blocking effect [11,12]. To obtain a high global $J_c$ of the bulk sample, it is very important to investigate the grain boundary blocking effect of this system. It has been founded that polycrystalline Sm-1111 are composed of Sm-1111 grains and a little of amorphous Fe-As phases (a-FeAs) [13-15]. They randomly co-exist and the a-FeAs is presumable cause of the grain boundary blocking effect.

In the case of polycrystalline Y-123 cuparates, an exponential decrease of global $J_c$ has also been observed at the grain boundary, as the grain misorientation angle increases above 3°-5° [16-20]. This decrease causes the intrinsic short superconducting order parameter of superconducting cuparates. On the other hand, in the case of MgB$_2$, superconducting order parameter has not been a serious problem for the decrease of global $J_c$. However, the impurity phase of MgO covering a part of the MgB$_2$ grains significantly prevents superconducting current between the MgB$_2$ grains [21-23].

In the 1111 system, short superconducting order parameter and/or impurity phase of a-FeAs are considered as the reason of decreasing global $J_c$ at the grain boundary. As is the case with Y-123 cuparates, the order parameter of 1111 system is also short.

However, global $J_c$ of polycrystalline Sm-1111 is much higher than that of polycrystalline Y-123 at the same temperature and magnetic field [14]. Therefore, a short order parameter of Sm-1111 may be not so much serious problem as that of Y-123 cuparates. There is room for further research into the enhancement of global $J_c$ for bulk Sm-1111.

One of the methods for the enhancement of $J_c$ is addition of metal [24-26]. For instance, Z. Gao et al reported the increase of transport $J_c$ for Sn addition in the Sr-122 system of iron based superconductors, and they suggest Sn addition strengthens the inter grain coupling [24]. On the other hand, the effective result of metal addition in 1111 system has not been reported. In this study, we focused on Sm-1111 with metal addition to obtain a high critical current density.

EXPERIMENTAL

The specifications of four kinds of Sm-1111 sample are listed in Table 1. To prepare these samples, three kinds of precursors were made by following way. Powder of Sm(99.9%), Fe (99.9%) and As (99.999%) were mixed and heated at 850 °C for 10 h in an evacuated quarts tube. The atomic ratio of the obtained precursors are the following (Sm:Fe:As = 1:3:3, 2:3:3 and 1:1:1), and they are named 133, 233, 111 powder derived from their elemental ratio respectively. For sample preparation, $SmFeAsO_{0.9}F_{0.1}$ was synthesized as sample 1 by a solid state reaction method [27]. Stoichiometric $Sm_2O_3$, $SmF_3$, 133 and 233 powders were ground in the mortar, compressed into pellets and sintered at 1200 °C for 40 h in an evacuated quarts tube. To obtain excess fluorine doped Sm-1111 as a sample 2, the synthesis path of our previous study was adopted [26]. 5 mol% of $SmF_3$ and 233 powder are added to the powder of Sample 1, and they were ground in the mortar, compressed into pellets and sintered at 900 °C for 40 h in the evacuated quarts tube. The sintered condition of sample 3 and 4 are same as that of sample 2. However, the starting materials are slightly different. Sample 3 is made of sample 1, $SmF_3$, 233 powder and indium (10 wt%). Sample 4 is made of sample 1, $InF_3$, 111 powder and indium (10 wt%).

The obtained polycrystalline samples were characterized by X-ray diffraction (XRD; Rigaku Rint 2500) using Cu Kα radiation. The polished surfaces of $SmFeAsO_{1-x}F_x$ were observed by scanning electron microscope (SEM; HITACHI SU-70) and optical microscope. The elemental compositions of samples were investigated by using energy dispersive x-ray spectrometry (EDX). The resistivity of the polycrystalline bulk samples were measured by the standard four-probe technique using Au electrodes. To estimate the magnetic $J_c$ by using the extended Bean model, magnetic hysteresis was measured by a superconducting quantum interference device (SQUID) magnetometer.

RESULTS

Figure 1 shows the XRD pattern for each sample. Almost all of the diffraction peaks are assigned to the Sm-1111 phase. However, minor peaks of SmOF are detected in the all samples except for sample 1. It is because an excess fluorine doping is performed in these samples. In sample 3, although the starting materials contain indium, diffraction peaks of indium are not detected. Indium is easy to move from inside to outside of the sample during a heat treatment, because of its low melting point. Actually, after sintering, indium is attached on the internal surface of the quartz tube. On the other hand, sample 4 shows the diffraction peals of indium.

Figure 2 (a) and (b) display the polished surface of a SEM image for each sample and EDX mapping for sample 1. The result of EDX mapping shows the different elemental compositions between grains and their interspace. From the quantitative analysis of EDX, the atomic ratio of grains are almost the same as that of Sm-1111, and interspaces are mainly composed of iron and arsenic. The interspaces are made of amorphous compounds, because the XRD diffraction peaks of the crystalline compounds made of Fe and As are not detected as shown in figure 1. Therefore, these grains and interspaces are recognized as Sm-1111 and a-FeAs respectively [13]. In sample 1, white line denotes the areas of a-FeAs and Sm-1111. These two areas randomly co-exist, and this inhomogeneous structure is formed in all region of the sample 1 and 2. On the other hand, a very homogeneous Sm-1111 area is observed in the indium added samples, and these images seems to exhibit the decrease of a-FeAs area by addition of indium, however, as shown in the insert of sample 4, the large areas of a-FeAs are also observed in other position of the same sample.

Figure 3 displays the optical photographs of polished surface for each sample. The different grain colors correspond to the different crystal faces from EBSD measurement [15]. In the case of the SEM image, it is difficult to confirm the grain size and the form in the sample 3 and 4 as shown in figure 2 (a). However, by using the optical photographs, grain sizes, forms and their boundaries can be recognized. Many Sm-1111 grains are packed without a-FeAs area in the indium added samples.

Figure 4 shows the temperature dependence of resistivity for the each sample.

Superconducting transition temperatures of $T_c^{onset}$ and $T_c^{zero}$ are given in Table 1. The sample 1 shows the lowest superconducting transition at 53.6 K, and gradual slope of resistivity is observed at the superconducting transition. This behavior suggests compositional inhomogeneity of sample 1. In sample 2, a steep slope of resistivity on the superconducting transition are observed, however, the resistivity of normal state is higher than that of sample 1. Meanwhile, indium added samples show the higher $T_c^{onset}$ and $T_c^{zero}$ and lower resistivity than sample 1 and 2, as shown in Table 1.

Figure 5 shows the magnetic field dependence of magnetic critical current density derived from the hysteresis loop width using the extended Bean model $J_c = 20\_m/Va(1-a/3b)$ for the all bulk samples taking the full sample dimensions of $1.0 \times 1.2 \times 6.0$ mm$^3$. As shown in figure 5, magnetic $J_c$ of sample 2 is higher than that of sample 1, and the indium added sample shows a much higher magnetic $J_c$ than sample 2. In sample 4, a value of magnetic $J_c$ at 25 kA/cm$^2$ was achieved. An obvious increase of magnetic $J_c$ is observed for the metal addition effect. These values of magnetic $J_c$ under self-magnetic field at 4.2 K are listed in the table 1.

DISSCUSSION

In this study, we succeeded in the enhancement of the magnetic critical current density of bulk Sm-1111 by the addition of indium. It is important to discuss the reason for this enhancement.

Comparing sample 1 with sample 2, the XRD diffraction peaks of sample 2 shifts to a higher angle and the impurity phase of SmOF is detected because of excess fluorine doping. In addition, the higher magnetic $J_c$ and rapid drop of resistivity are observed in the sample 2. It means the fluorine concentration of sample 2 is higher and more homogeneous than that of sample 1. However, normal state resistivity of sample 2 is much higher than that of sample 1, and a-FeAs area of sample 2 seems to be larger than that of sample 1 in the figure 2 (a). The increase of normal state resistivity causes the increase of a-FeAs area, because of the higher resistivity of a-FeAs than that of Sm-1111.

Comparing sample 2 with sample 3, the higher magnetic $J_c$ and lower normal state resistivity are observed in sample 3. We suggest the effect of indium addition for polycrystalline Sm-1111 is the removal of a-FeAs areas from interfaces of Sm-1111 grains and the bringing together the Sm-1111 grains. For instance, in sample 1 and 2 of the figure 2 (a), a-FeAs areas covered the Sm-1111 grains are observed all around the entire sample, on the other hand, a-FeAs areas are located partially in the indium added samples. It means that the total contact surfaces of Sm-1111 grains are increased by the indium addition. This increase contributes the enhancement of magnetic $J_c$ and the decrease of normal state resistivity. Interestingly, there are no diffraction peaks of elemental indium in sample 3. EDX measurement also did not show any indium signals. Therefore, all of the added indium moves onto the outside of sample 3. In the moving of indium from inside to outside of sample, the effect of indium addition described above may occur.

Comparing sample 3 with sample 4, sample 4 shows a much higher magnetic critical current density and lower normal state resistivity. The diffraction peaks of indium are detected in the sample 4. EDX result also shows the existence of indium, and it is partly located between the grains. The actual amount of indium is about 5 wt%. The detected indium is mainly derived from InF$_3$ because of the high melting point of this material. As is the case with sample 3, a-FeAs areas are removed from interfaces of Sm-1111 grains, and Sm-1111 grains bring together in the sample 4. This indium addition effect of sample 4 seems to be more developed than that of sample 3. As shown in figure 2, a-FeAs areas are hardly observed in the sample 4. We suggest the reason of increasing the magnetic $J_c$ is increase of total contact surfaces of Sm-1111 grains.

CONCLUSION

We succeeded in the enhancement of magnetic critical current density by indium addition. The value of magnetic $J_c$ is around $2.5 \times 10^4$ A/cm$^2$ at 4.2 K under self field in sample 4. One of the reasons why the enhancement of magnetic $J_c$ is observed is the increase of total contact surface between Sm-1111 grains. We suggest the effect of indium addition is the removal of the a-FeAs, which covers the Sm-1111 grains, from the interface of Sm-1111 grains and the bringing together of Sm-1111 grains.

ACKNOWLEDGEMENTS
This work was supported in part by the Japan Society for the Promotion of Science through 'Funding program for World-Leading Innovative R&D on Science Technology Program (FIRST)' and Japan Science and Technology Agency through Strategic International Collaborative Research Program (SICORP-EU-Japan).

Figure 1

(Color online) Powder XRD pattern for each sample. black bars at the bottom show the calculated Bragg diffraction positions of SmFeAsO$_{0.92}$F$_{0.08}$. The arrows denote peaks of SmOF and In, respectively.

Figure 2

(Color online) (a): SEM images of the polished surface of samples. Inserts are the expanded view of sample 1 and 4. (b): EDX mapping of inserted image for sample 1. Blue, purple, yellow and red signals denote the concentration of Fe, As, Sm and O, respectively.

Figure 3

(Color online) Optical micrograph of polished surface for the each sample. Scale bar at the bottom right corner is 10 μm.

Figure 4

(Color online) Resistivity $\rho$ versus temperature $T$ for the each sample. The black square, red circle, green triangle and blue inverted square denote the sample 1, 2, 3 and 4, respectively. Insert shows the expanded view.

Figure 5

(Color online) Mgnetic $J_c$ versus magnetic field $\mu_0H$ for the each sample. The black square, red circle, green triangle and blue inverted square denote the sample 1, 2, 3 and 4, respectively.

Table 1. Sample specifications. The values of magnetic Jc are measured at 4.2 K under self-magnetic field.

| Sample no. | Raw materials | $T_c^{onset}$ (K) | $T_c^{zero}$ (K) | $J_c^{magnetic}$ (A/cm$^2$) |
|---|---|---|---|---|
| Sample 1 | Sm$_2$O$_3$, SmF$_3$, 133, 233 powder | 53.6 | 31.3 | 14979 |
| Sample 2 | sample1, SmF$_3$ 233 powder | 54.8 | 42.5 | 17019 |
| Sample 3 | sample1, SmF$_3$ 233 powder, In | 55.6 | 44.4 | 20026 |
| Sample 4 | sample1, InF$_3$ 111 powder, In | 54.8 | 44.4 | 24728 |

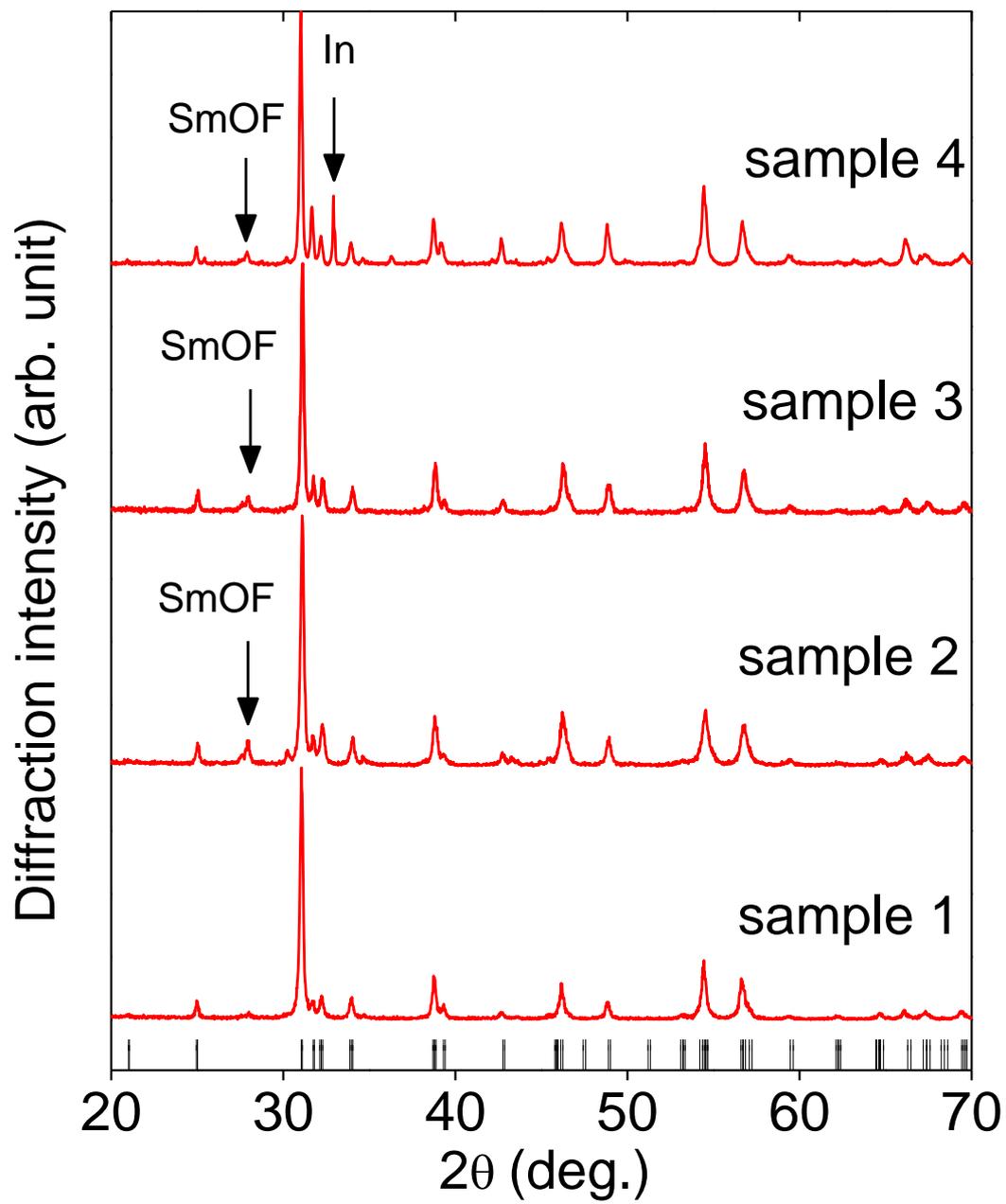

Fig. 1

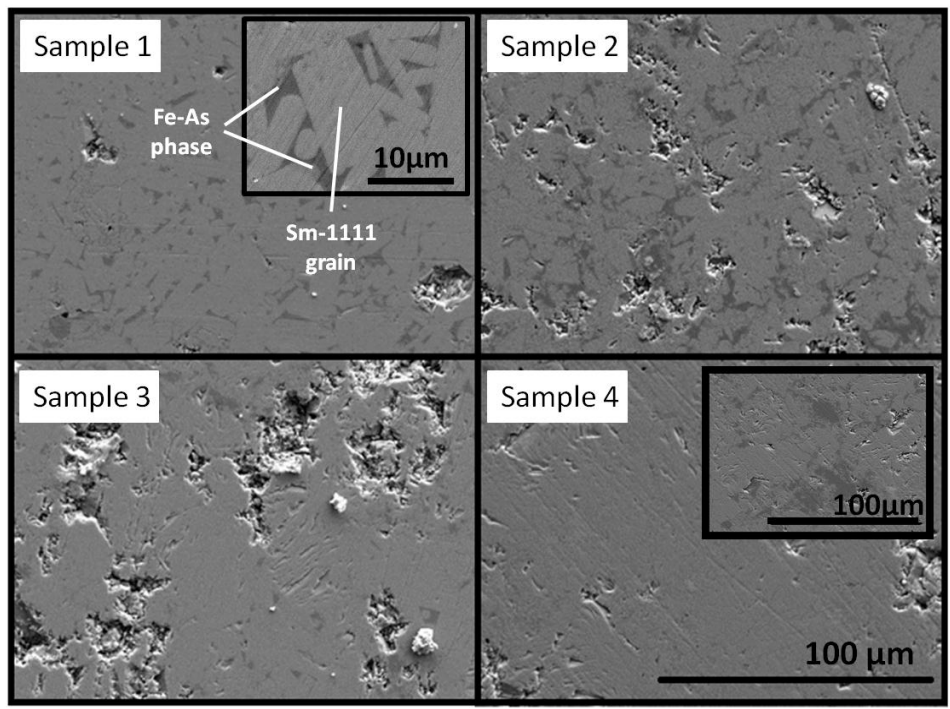

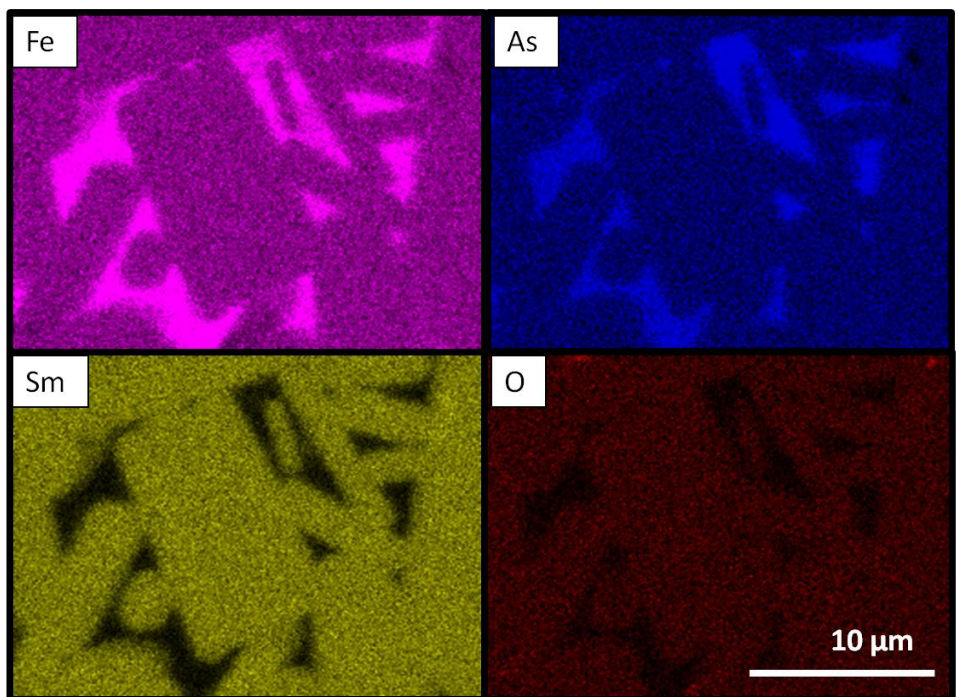

Fig. 2

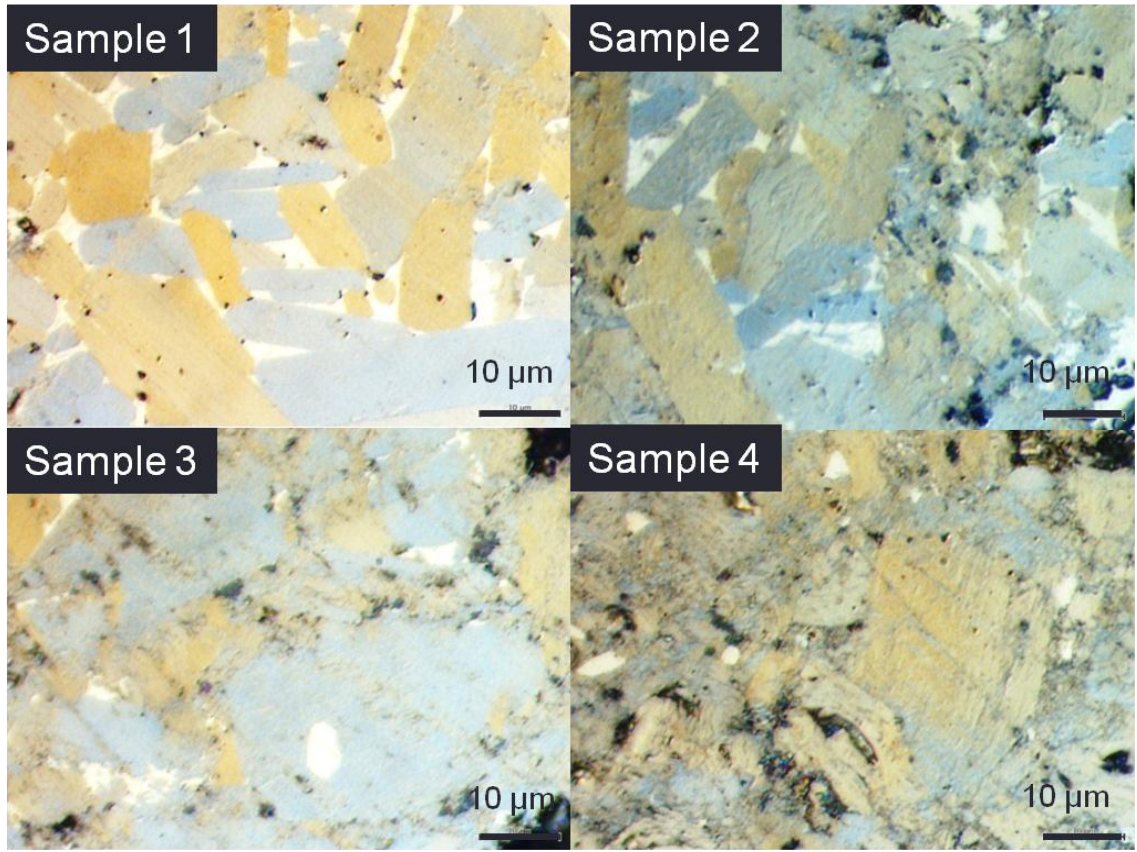

Fig. 3

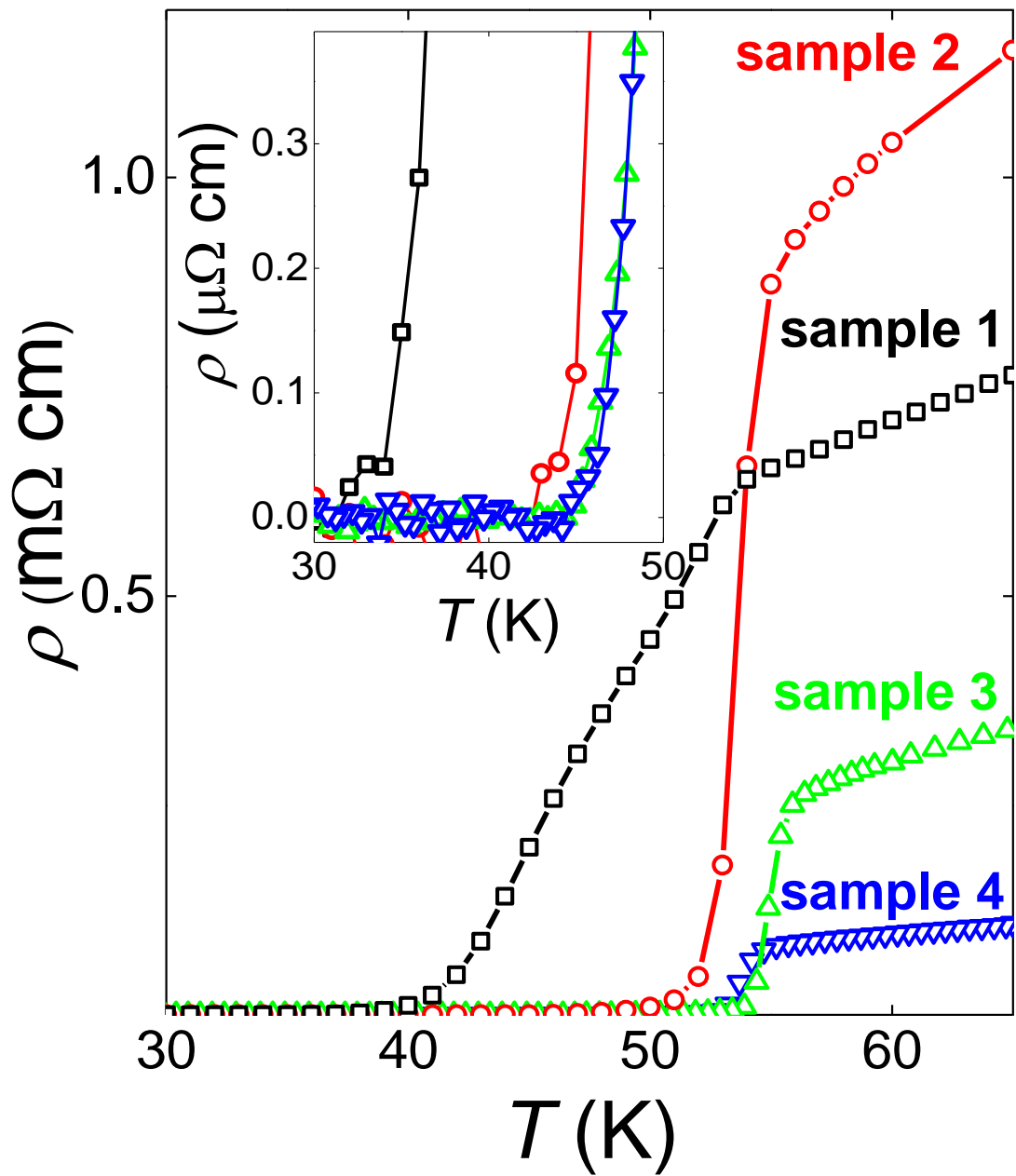

Fig. 4

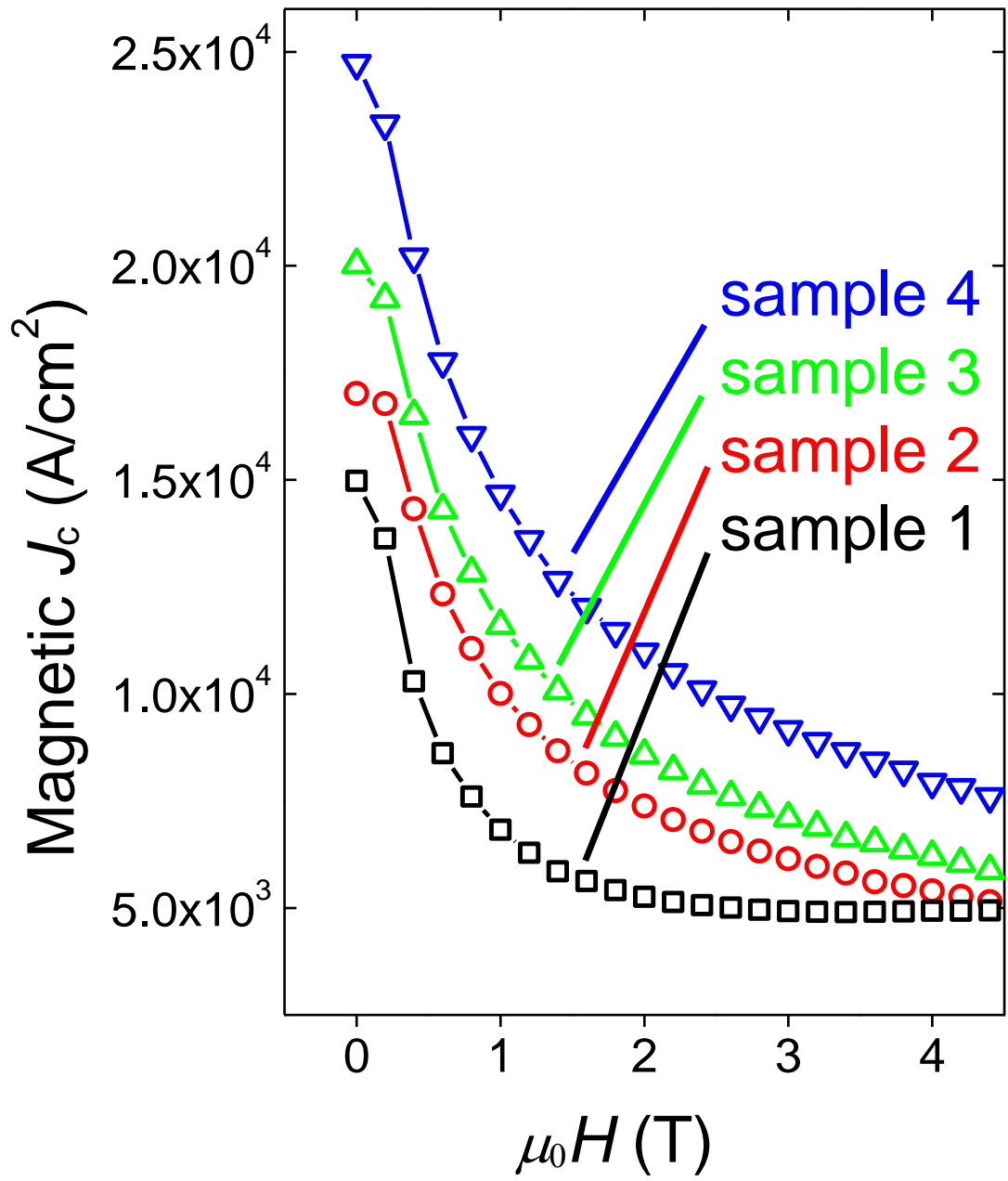

Fig. 5